\documentclass[useAMS,usenatbib]{mn2e} 
\usepackage{amsmath,natbib,graphics}
\usepackage{epsfig,subfigure}
\usepackage{amssymb}
\let\la=\lesssim    
\def\reff@jnl#1{{\rm#1\/}}
\def\aj{\reff@jnl{AJ}}                  % Astronomical Journal
\def\araa{\reff@jnl{ARA\&A}}            % Annual Review of Astron and Astrophys
\def\apj{\reff@jnl{ApJ}}                % Astrophysical Journal
\def\apjl{\reff@jnl{ApJ}}               % Astrophysical Journal, Letters
\def\apjs{\reff@jnl{ApJS}}              % Astrophysical Journal, Supplement
\def\ao{\reff@jnl{Appl.Optics}}         % Applied Optics
\def\apss{\reff@jnl{Ap\&SS}}            % Astrophysics and Space Science
\def\aap{\reff@jnl{A\&A}}               % Astronomy and Astrophysics
\def\aapr{\reff@jnl{A\&A~Rev.}}         % Astronomy and Astrophysics Reviews
\def\aaps{\reff@jnl{A\&AS}}             % Astronomy and Astrophysics, Supplement
\def\azh{\reff@jnl{AZh}}                % Astronomicheskii Zhurnal
\def\baas{\reff@jnl{BAAS}}              % Bulletin of the AAS
\def\jrasc{\reff@jnl{JRASC}}            % Journal of the RAS of Canada
\def\memras{\reff@jnl{MmRAS}}           % Memoirs of the RAS
\def\mnras{\reff@jnl{MNRAS}}            % Monthly Notices of the RAS
\def\pra{\reff@jnl{Phys.Rev.A}}         % Physical Review A: General Physics
\def\prb{\reff@jnl{Phys.Rev.B}}         % Physical Review B: Solid State
\def\prc{\reff@jnl{Phys.Rev.C}}         % Physical Review C
\def\prd{\reff@jnl{Phys.Rev.D}}         % Physical Review D
\def\prl{\reff@jnl{Phys.Rev.Lett}}      % Physical Review Letters
\def\pasp{\reff@jnl{PASP}}              % Publications of the ASP
\def\pasj{\reff@jnl{PASJ}}              % Publications of the ASJ
\def\qjras{\reff@jnl{QJRAS}}            % Quarterly Journal of the RAS
\def\skytel{\reff@jnl{S\&T}}            % Sky and Telescope
\def\solphys{\reff@jnl{Solar~Phys.}}    % Solar Physics
\def\sovast{\reff@jnl{Soviet~Ast.}}     % Soviet Astronomy
\def\ssr{\reff@jnl{Space~Sci.Rev.}}     % Space Science Reviews
\def\zap{\reff@jnl{ZAp}}                % Zeitschrift fuer Astrophysik
\def\nat{\reff@jnl{Nature}}             % Nature 

\newcommand{\iras}{{\em IRAS~}}
\newcommand{\uv}{$(u,v)$}

\title[Dust-correlated cm-wavelength continuum emission from
translucent clouds]{Dust-correlated cm-wavelength continuum emission
  on translucent clouds $\zeta$~Oph and LDN~1780}

\author[Vidal et al.]{M.~Vidal$^{1,2}$\thanks{E-mail: mvidal@jb.man.ac.uk
    (MV)}, S.~Casassus$^{2,3}$, C.~Dickinson$^{1}$, A.~N.~Witt$^{4}$,
  P.~Castellanos$^{2}$,\newauthor R.~D.~Davies$^{1}$, R.~J.~Davis$^{1}$,
  G.~Cabrera$^{2}$, K.~Cleary$^{8}$, J.~R.~Allison$^{5}$, \newauthor
  J.~R.~Bond$^{6}$, L.~Bronfman$^{2}$, R.~Bustos$^{2,7}$, M.~E.~Jones$^{5}$,
  R.~Paladini$^{9}$, \newauthor T.~J.~Pearson$^{8}$,
  A.~C.~S.~Readhead$^{8}$, R.~Reeves$^{10}$,
  J.~L.~Sievers$^{6}$,  A.~C.~Taylor$^{5}$.\\
  $^{1}$ Jodrell Bank Centre for Astrophysics, Alan Turing Building,
  School of Physics and Astronomy, The University of Manchester,\\
  Oxford Road, Manchester M13 9PL, UK. \\
  $^{2}$ Departamento de Astronom\'{\i}a, Universidad de Chile,
  Casilla 36-D, Santiago, Chile.\\
  $^{3}$ Observatoire de Paris, LUTH and Universit\'e Denis Diderot, Place
  J. Janssen, 92190 Meudon,
  France.  \\
  $^{4}$ Ritter Astrophysical Research Center, University of
  Toledo, Toledo, OH 43606, USA.\\
  $^{5}$ Oxford Astrophysics, University of Oxford, Denys
  Wilkinson Building, Keble Road, Oxford, OX1 3RH, UK.\\
  $^{6}$ Canadian Institute for Theoretical Astrophysics,
  University of Toronto, Toronto, Canada.\\
  $^{7}$ Departamento de Astronom\'{i}a, Universidad de
  Concepci\'{o}n, Casilla 160-C, Concepci\'{o}n, Chile.\\
  $^{8}$ Cahill Center for Astronomy and Astrophysics, Mail Code
  \mbox{249-17}, California Institute of Technology, Pasadena, CA
  91125, USA.\\
  $^{9}$ Spitzer Science Center, 1200 East California
  Boulevard, Pasadena, CA 91125, USA\\
  $^{10}$ Departamento de Ingenier\'{i}a El\'{e}ctrica, Universidad de
  Concepci\'{o}n, Concepci\'{o}n, Chile }

\begin{document}

%\date{Accepted 2011 February 18. Received 1988 December 14; in
%  original form 1988 October 11}

%\pagerange{\pageref{firstpage}--\pageref{lastpage}} \pubyear{2002}

\maketitle

\label{firstpage}

\begin{abstract}
  The diffuse cm-wave IR-correlated signal, the ``anomalous'' CMB foreground,
  is thought to arise in the dust in cirrus clouds. We present Cosmic
  Background Imager (CBI) cm-wave data of two translucent clouds, $\zeta$~Oph
  and LDN~1780 with the aim of characterising the anomalous emission in the
  translucent cloud environment.

  In $\zeta$~Oph, the measured brightness at 31~GHz is $2.4\sigma$ higher than
  an extrapolation from 5~GHz measurements assuming a free-free spectrum on
  8~arcmin scales. The SED of this cloud on angular scales of 1$^\circ$ is
  dominated by free-free emission in the cm-range. In LDN~1780 we detected a
  3~$\sigma$ excess in the SED on angular scales of 1$^{\circ}$ that can be
  fitted using a spinning dust model. In this cloud, there is a spatial
  correlation between the CBI data and IR images, which trace dust. The
  correlation is better with near-IR templates (\iras 12 and 25~$\mu$m) than
  with \iras 100~$\mu$m, which suggests a very small grain origin for the
  emission at 31~GHz.
 
  We calculated the 31~GHz emissivities in both clouds. They are similar and
  have intermediate values between that of cirrus clouds and dark
  clouds. Nevertheless, we found an indication of an inverse relationship
  between emissivity and column density, which further supports the VSGs
  origin for the cm-emission since the proportion of big relative to small
  grains is smaller in diffuse clouds.

\end{abstract}


%%%%%%%%%%%%%%%%%%%%%%%%%%%%%%%%%%%%%%%%%%%%%%%%%%%%%%%%%%%%%%%%%%%%%%
%%%%%%%%%%%%%%%%%%%%%%%%%%%%%%%%%%%%%%%%%%%%%%%%%%%%%%%%%%%%%%%%%%%%%%

\section{Introduction}

Amongst the challenges involved in the study of Cosmic Microwave Background
(CMB) anisotropies is the subtraction of Galactic foreground emission. The
detailed study of these foregrounds led to the discovery of a new
radio-continuum mechanism in the diffuse interstellar medium (ISM).  It was
first detected by the {\em Cosmic Background Explorer} as a diffuse all-sky
dust-correlated signal with a flat spectral index between 31 and 53~GHz
\citep{kogut96a,kogut96b}. With ground based observations, \citet{leitch97}
observed high Galactic latitude clouds and detected an excess at 14.5~GHz
that, because of a lack of H$\alpha$ emission, could not be accounted for by
free-free with typical temperatures ($T\sim 10^4$K).

\begin{figure*}  % fig 1
  \centering \includegraphics[width=1\textwidth]{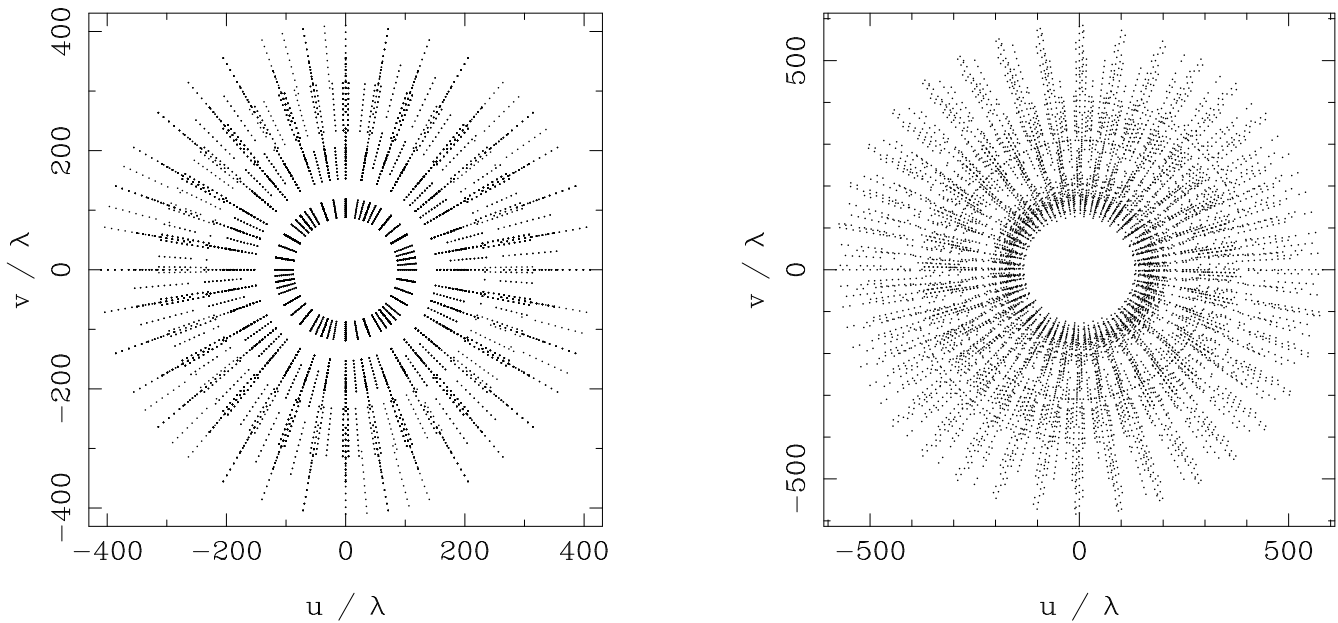}
  \caption{($u,v$) coverage of the CBI (left) and CBI2 (right) in the
    configurations used for the observations of $\zeta$ Oph on the left and
    LDN~1780 on the right. Axes are in units of wavelength ($\lambda$).}
  \label{uvplane}
\end{figure*}

Different emission mechanisms have been proposed for the anomalous emission,
such as spinning dust \citep{DL98a, DL98b}, magnetic dust \citep{DL99}, hot
(T$\sim$10$^6$~K) free-free \citep{leitch97} and hard synchrotron radiation
\citep{bennett03}. To date, the evidence favours the spinning dust grain model
\citep{finkbeiner04b, oliveira04, watson05, casassus1622, casassus08,
  castellanos10} in which very small grains (VSGs) with a non-zero dipole
moment rotating at GHz frequencies emit cm-wave radiation. The emission has
its peak at $\sim$20-40~GHz and the model predicts that it is dominated by the
smallest grains, possibly polycyclic aromatic hydrocarbons
(PAHs). \citet{ysard10} found a correlation across the whole sky using {\em
  WMAP} data between 23~GHz maps and \iras 12~$\mu$m, which supports the VSG
origin for the cm-wave emission.

\begin{figure*}  % fig 2
  \centering \includegraphics[width=1\textwidth,angle=0]{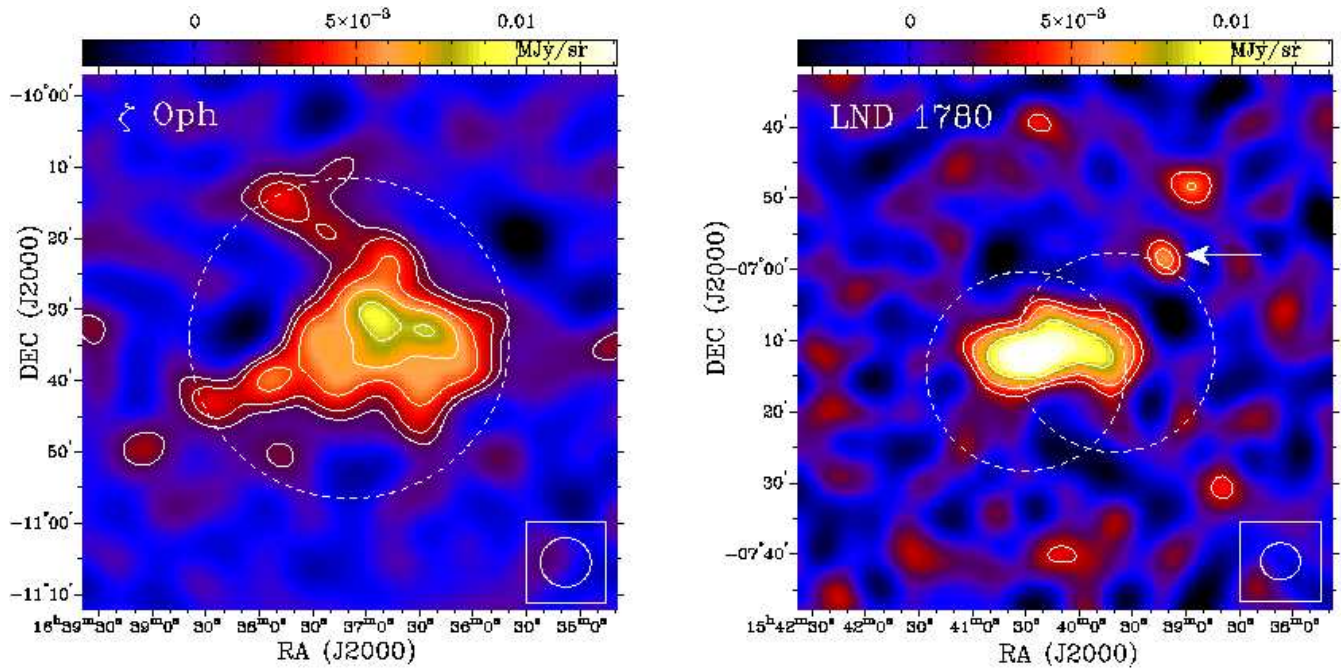}
  \caption{ CLEANed 31~GHz images of $\zeta$ Oph (left) and LDN~1780
    (right). Contours are 10, 30, 50, 70, 90\% of the peak brightness, which
    is $9\times 10^{-3}$ MJy~sr$^{-1}$ in $\zeta$ Oph and 0.016 MJy~sr$^{-1}$
    in LDN~1780. The FWHM primary beams of CBI and CBI2 are shown as dashed
    lines (in LDN~1780, there are two pointings). These images have not been
    beam-corrected. The point source NVSS~153909-065843 in LDN~1780 is marked
    with an arrow. The synthesised beam is shown at the bottom right corner
    of each frame.}
  \label{fig:restored}
\end{figure*}
The radio-IR correlation suggests that the ``cirrus'' clouds are responsible
for the anomalous emission at high Galactic latitudes.  Cirrus is the
large-scale filamentary structure detected by {\em IRAS} \citep{low84}. It is
seen predominantly at 60 and 100~$\mu$m in the {\em IRAS} bands and the origin
of the IR-radiation is generally ascribed to dust continuum emission with a
contribution from atomic lines \citep[e.g. O\,{\sc
  i}~63~$\mu$m,][]{stark90}. On shorter wavelengths (e.g. 12~$\mu$m), the
emission is from polycyclic aromatic hydrocarbons (PAHs) subject to thermal
fluctuations. Cirrus clouds span a wide range of physical parameters; most are
atomic and some are molecular \citep[e.g.][]{miville02,
  snow06}. \citet{stark94}, using a combination of absorption and emission
line measurements, found H{\sc i} gas temperatures in the range between 20 and
$>$~350~K. The low temperatures correspond to the coldest clumps in the clouds
that represent the cores of the much more widely distributed and hotter H{\sc
  i} gas. Cirrus clouds are pervaded by the interstellar radiation field
(IRF).  They have column densities $N$(H{\sc i})$ \approx 1-10 \times 10^{20}$
cm$^{-2}$, which correspond to visual extinctions of $A_V \leq 1$~mag assuming
a typical dust-to-gas ratio of 100. It is commonly found that these clouds are
not gravitationally bound and that their kinematics are dominated by
turbulence \citep{magnani85}.  Cirrus clouds are difficult to characterise
because of their low column densities, and are thus not the best place to
study the anomalous emission.

Local known clouds offer an opportunity to characterise the cm-emission. A
number of investigations have been made of on these clouds: the Perseus
molecular cloud \citep{watson05}, $\rho$~Oph \citep{casassus08}, RCW~175
\citep{dickinson09}, LDN1111 \citep{scaife09}, M78 \citep{castellanos10},
among others.  \citet{finkbeiner04a} using {\em WMAP} (23-94~GHz) and Green
Bank (5,8,10~GHz) data fitted a spinning dust model to the emission from the
cloud LDN1622. \citet{casassus1622}, with interferometric observations at
31~GHz of the same cloud, detected bright cm-emission where no emission from
any known mechanism was expected. They performed a morphological analysis and
found a better cross-correlation between 31~GHz and \iras~12~$\mu$m than with
\iras~100~$\mu$m. \citet{watson05} with the COSMOSOMAS experiment discovered
strong emission in the frequency range 11-17~GHz in the Perseus Molecular
Cloud (G159.6-18.5). In this cloud, the spectral energy distribution (SED) is
well-fitted by a spinning dust model. A disadvantage of these studies, in the
aim of characterising anomalous emission as a foreground to the CMB, is that
the clouds already studied potentially constitute a very different phase of
the ISM than the cirrus, so inferences drawn from them may not be applicable
to understanding the anomalous emission from the cirrus seen at high Galactic
latitudes.

In this paper we present cm-wave continuum data of the translucent clouds
$\zeta$~Oph and LDN~1780 acquired with the Cosmic Background Imager
(CBI). Translucent clouds are interstellar clouds with some protection from
the radiation field in that their extinction is in the range $A_V\sim1-4$~mag
\citep{snow06}. They can be understood as photo-dissociation-regions
(PDR). Across translucent clouds, carbon undergoes a transition from
singly-ionised into neutral atomic or molecular (CO) form. In translucent
clouds, physical properties such as density and temperature, and environmental
conditions, such as exposure to the IRF, are intermediate between those of
dense clouds and those of transparent (cirrus) clouds. By bridging the gap in
physical conditions, translucent clouds are test beds for the extrapolation of
the radio/IR relative emissivities seen in dense clouds.

The Lynds Dark Nebula (LDN)~1780 is a high Galactic latitude ({\em l} =
359.0$^{\circ}$, {\em b} = 36.7$^{\circ}$) translucent region at a distance of
110$\pm$10~pc \citep{franco89}. \citet{ridd06} found that the spatial
distribution of the mid-IR emission differs significantly from the emission in
the far-IR. Also, they show using IR colour ratios that there is an
overabundance of PAHs and VSGs with respect to the solar neighbourhood
\citep[as tabulated in][]{boulanger88}, although their result can be explained
by an IRF that is overabundant in UV photons compared to the standard IRF
\citep{witt10}. Using an optical-depth map constructed from {\em
  ISO}~200~$\mu$m emission, \citet{ridd06} found a mass of $\sim$
18~M$_{\odot}$ and reported no young stellar objects based on the absence of
colour excess in point sources. Because of the morphological differences in
the IR, this cloud is an interesting target to make a morphological comparison
with the radio data, in order to determine the origin of the anomalous
emission. The free-free emission from this cloud is very low, which is
favourable for a study of the radio-IR correlated emission.

The cloud coincident with $\zeta$~Oph is a prototypical and well-studied
translucent cloud. $\zeta$~Oph itself is an O9Vb star at a distance of 140$\pm
16$~pc \citep{hipparcos97}. In this line of sight, the total H-nucleus column
density is fairly well determined, $N\text{(H)} \sim 1.4 \times 10^{21}
\text{cm}^{-2}$, with 56\% of the nuclei in molecular form
\citep{morton75}. Although the observations reveal several interstellar
components at different heliocentric velocities, the one at $v{_\odot} =
-14.4$~km s$^{-1}$ contains most of the material and is referred to as the
$\zeta$~Oph cloud. There have been numerous efforts to build chemical models
of this cloud \citep{BD77,vDB86, viala88}. In the pioneering work of
\citet{BD77}, a two shell model is proposed to fit the observational data: a
cold and denser core surrounded by a diffuse envelope. The density and
temperature in these models are in the range $n=250-2500$ cm$^{-3}$, $T =
20-100$~K for the core and $n=200-500$ cm$^{-3}$, $T = 100-200$~K for the
envelope. These conditions approach those of the cirrus cloud cores whose
densities lie in the same range \citep{turner94}.

The rest of the paper is organised as follows: in \textsection~2 we describe
the data acquisition, image reconstruction and list the auxiliary data used
for comparison with the CBI data. \textsection~3 contains a spectral and
morphological analysis of both clouds. In \textsection~4 there is a comparison
of the 31~GHz emissivity of dark, translucent and cirrus
clouds. \textsection~5 presents our conclusions.

%%%%%%%%%%%%%%%%%%%%%%%%%%%%%%%%%%%%%%%%%%%%%%%%%%%%%%%%%%%%%%%%%%%%%%
%%%%%%%%%%%%%%%%%%%%%%%%%%%%%%%%%%%%%%%%%%%%%%%%%%%%%%%%%%%%%%%%%%%%%%

\section{Data}

\begin{figure*} % fig 3
  \centering
  \includegraphics[width=1\textwidth, angle=0]{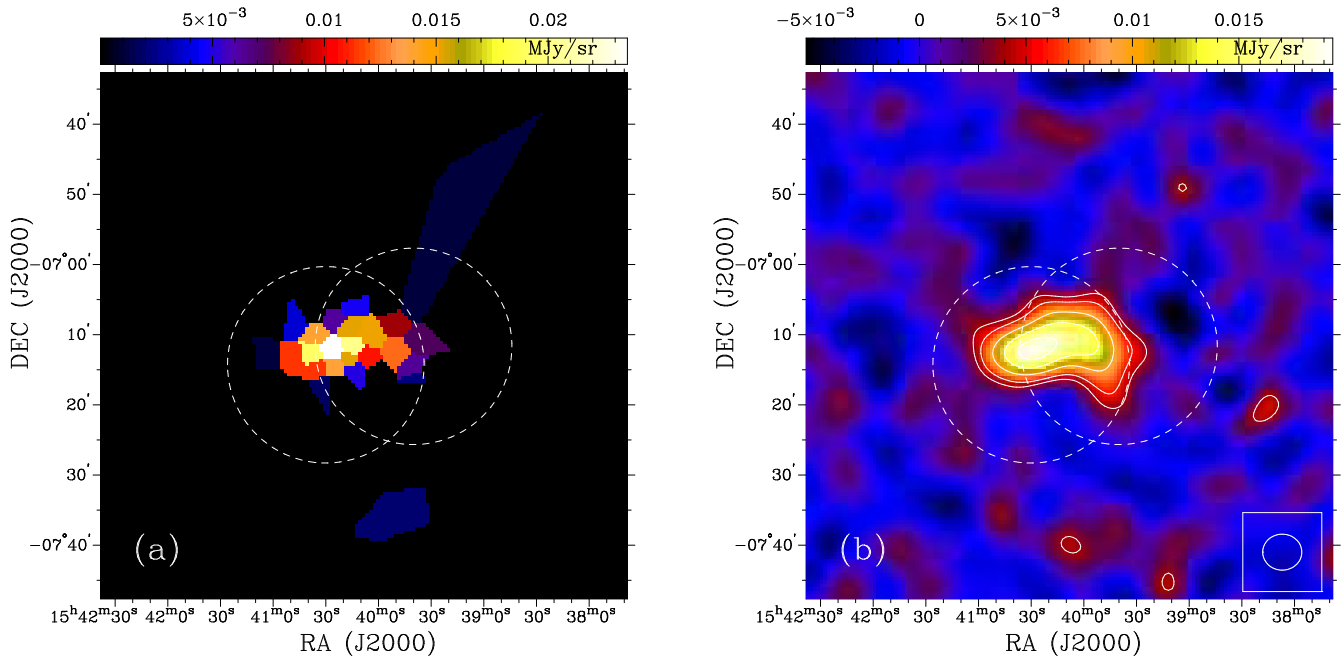}
  \caption{ Left: Voronoi model of the visibilities of LDN 1780. This model
    has 55 polygons.  Right: the convolution of the model in the left with the
    synthetic beam of the data, obtained using natural weights.}
  \label{fig:vor_1780}
\end{figure*}

\subsection{31 GHz Observations}

The observations at 31~GHz were carried out with the Cosmic Background Imager
\citep[CBI1,][]{padin02}, a 13 element interferometer located at an altitude
of 5000~m in the Chajnantor plateau in Chile. Each antenna is 0.9~m in
diameter and the whole array is mounted on a tracking platform, which rotates
in parallactic angle to provide uniform $uv$ coverage. The primary beam has
full-width half maximum (FWHM) of 45.2~arcmin at 31~GHz and the angular
resolution is $\sim$8~arcmin. The receivers operate in ten frequency channels
from 26 to 36~GHz. Each receiver measures either left (L) or right (R)
circular polarisation. The interferometer was upgraded during 2006-2007 with
1.4~m dishes (CBI2) to increase temperature sensitivity (Taylor et al. in
prep.). The primary beam is 28.2~arcmin FWHM at 31~GHz and the angular
resolution was increased to $\sim$4~arcmin.

Table \ref{observations} summarises the observations. Both clouds were
observed in total intensity mode (all receivers measuring L only). $\zeta$~Oph
was observed using CBI1 in a single position while LDN~1780 was observed in
two different pointings: L1780E and L1780W by CBI2. The configurations of the
CBI1 and CBI2 interferometers result in the ($u,v$) coverage shown in Fig.
\ref{uvplane}.

We reduced the data using the same routines as those used for CMB data
analysis \citep{pearson03, readhead04a, readhead04b}. Integrations of 8-min on
source were accompanied by a trail field, with an offset of 8~arcmin in RA,
observed at the same hour angle for the subtraction of local correlated
emission (e.g. ground spillover). Flux calibration is tied to Jupiter
\citep[with a brightness temperature of $146.6\pm0.75$~K,][]{hill09}.

\begin{table}
  \centering
  \caption{Summary of observations. The coordinates are J2000 and the time
    quoted is the observation time on-source.}
  \begin{tabular}{lcccc}
    \hline
    \hline
    Source    & Date & R.A.  &  decl.      & Time   \\
    \hline
    $^1\zeta$~Oph &8/7/04 & $16^\textup{h}37^\textup{m}9^\textup{s}.5$& $-10^{\circ}34'01''$ & 8000~s  \\
    $^2$L1780E &13, 28/4/07 & $15^\textup{h}40^\textup{m}30^\textup{s}$ & $-07^{\circ}14'18''$ & 12000~s \\
    $^2$L1780W &17, 18/4/07 & $15^\textup{h}39^\textup{m}40^\textup{s}$ & $-07^{\circ}11'40''$ & 8000~s  \\
    \hline
  \end{tabular}  
   \begin{flushleft}
     $^1$: observed with CBI1\\
     $^2$: observed with CBI2
  \end{flushleft}  
  \label{observations}
\end{table}

{\small %%% TABLE 2
  \begin{table*}
    \centering
    \caption{Auxiliary data used.}
    \begin{tabular}{lccc}
      \hline 
      \hline 
      Survey/Telescope & $\lambda$/$\nu$ &
      $\theta^1$ & Reference\\
      \hline
      SHASSA  & 656.3~nm & 0.8~arcmin  & \citet{gaustad01}  \\
      {\em Spitzer}$^2$  & 8~$\mu$m & 2$''$ & \citet{irac}  \\
      {\em IRAS}/IRIS  & 12, 25, 60~\&~100~$\mu$m &  3.8 - 4.3~arcmin  &\citet{iris} \\
      {\em ISO}  & 100 \& 200~$\mu$m & $\sim 1$~arcmin   &  \citet{ISO}\\
      {\em COBE}/DIRBE  & 100, 140~\&~240~$\mu$m & $\sim 0.7^{\circ} $  & \citet{Hauser98} \\
      \hline     
      {\em WMAP}  & 23, 33, 41, 61,~\&~94~GHz & 53 - 13~arcmin  &\citet{wmap} \\
      PMN  & 4.85~GHz & $\sim5$~arcmin  &  \citet{PMN93} \\
      Stockert& 2.72~GHz & $\sim20$~arcmin  &  \citet{stockert11}  \\
      HartRAO& 2.326~GHz & $\sim20$~arcmin  &  \citet{rhodes2326}\\
      Stockert & 1.4~GHz & $\sim35$~arcmin  &  \citet{stockert21} \\
      Parkes& 0.408~GHz & 51~arcmin  & \citet{haslam81}  \\      
    \end{tabular}
    \begin{flushleft}
      $^1$: $\theta$ is the angular resolution FWHM \\
      $^2$: Spitzer program 40154
    \end{flushleft}
    \label{tab:auxdata}
  \end{table*}
}

\subsubsection{Image reconstruction}

The reduced and calibrated visibilities were imaged with the CLEAN algorithm
\citep{hogbom74} using the DIFMAP package \citep{difmap}.  We chose natural
weights in order to obtain a deeper restored image. The theoretical noise
level (using natural weights, as expected from the visibility weights
evaluated from the scatter of individual frames) is 4.9~mJy~beam$^{-1}$ for
$\zeta$~Oph. Fig. \ref{fig:restored} shows the CLEAN image of
$\zeta$~Oph. This image was then corrected by the primary beam response of
CBI1 at 31~GHz (45.2~arcmin).  For LDN~1780, we obtained CLEAN images of the two
fields. The estimated noise level is 4.3~mJy~beam$^{-1}$ for L1780E and
3.0~mJy~beam$^{-1}$ for L1780W. In LDN~1780W a point source,
NVSS~153909-065843 \citep{condon98}, lies close to the NNW sector of the
half-maximum contour of the CBI2 primary beam in Fig. \ref{fig:restored}. This
source allowed us to fix the astrometry of the data which corresponded to an
offset of 30$''$ and 56$''$ in R.A. and Dec. These shifts are consistent with
the root mean square (rms) pointing accuracy of $\sim$~0.5~arcmin. The restored
CBI2 fields of LDN~1780 were then combined into the weighted mosaic shown on
Fig. 2, according to the following formula: $ I_M(\mathbf{x})={\sum_p
  (w_p(\mathbf{x})I_p(\mathbf{x})/A_p(\mathbf{x}))}/{\sum_p w_p(\mathbf{x})},$
where $w_p(\mathbf{x})=A_p^2/\sigma_p^2$, with $p$ the label for the $p$th
pointing and $A_p(\mathbf{x})$ is the primary beam response. 

We also used an alternative method to restore the visibilities in order to
check the CLEAN reconstruction. We applied the Voronoi image reconstruction
(VIR) from \citet{cabrera08}, that is well suited to noisy data sets. This
novel technique uses a Voronoi tessellation instead of the usual grid, and has
the advantage that it is possible to use a smaller number of free parameters
during the reconstruction. Moreover, this technique provides the optimal image
in a Bayesian sense, so the final image is unique.

Fig. \ref{fig:vor_1780} shows the VIR image for LDN~1780. The point source was
subtracted from the visibilities before the reconstruction. The result is
visibly better because the image negatives are less pronounced than CLEAN
(Fig. \ref{fig:restored}). The largest negative intensity with CLEAN is
$-0.007$~MJy/sr whereas with VIR is $-0.005$~MJy/sr. The dynamic range is
larger with the VIR reconstruction: 14.8 versus 11.5 from CLEAN. The
disadvantage of the VIR reconstruction is the large amount of CPU time
required.

Although the VIR reconstruction is better, both techniques gives very similar
results. Because of this, we trust the CLEAN reconstruction and used those
images for the rest of the investigation.

\subsection{Auxiliary data}

\begin{figure*} % fig 4
  \centering
  \includegraphics[width=0.23\textwidth,angle=-90]{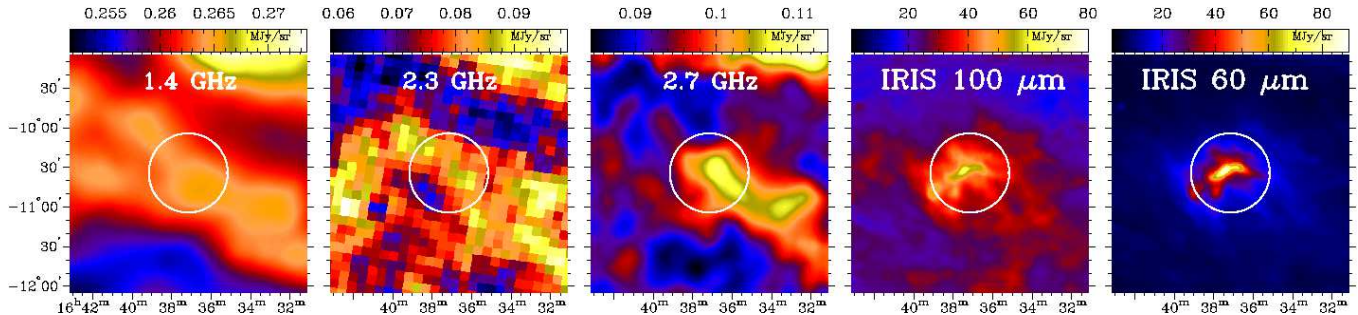}
  \caption{Maps from the auxiliary data around the extraction aperture for
    $\zeta$ Oph, 3$^{\circ}$ per side, in J2000 equatorial coordinates. The
    1$^{\circ}$ extraction aperture is shown.}
  \label{fig:mosaic_zoph}
\end{figure*}

\begin{figure*}  % fig 5
  \centering
  \includegraphics[width=.23\textwidth,angle=-90]{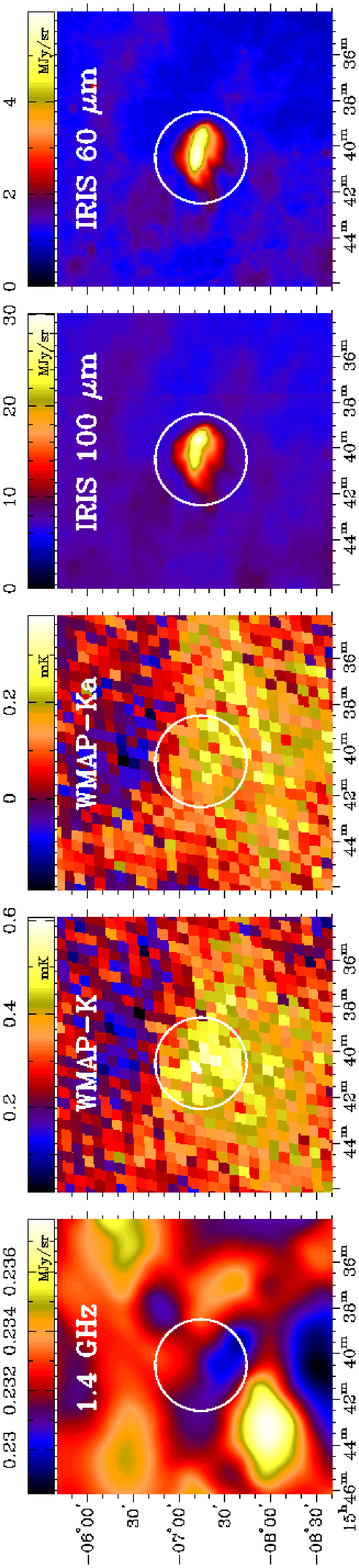}
  \caption{Maps from the auxiliary data around the extraction aperture for
    LDN~1780, 3$^{\circ}$ per side, in J2000 equatorial coordinates.  The
    1$^{\circ}$ extraction aperture is shown.}
  \label{fig:mosaic_LDN1780}
\end{figure*}

Table \ref{tab:auxdata} lists the auxiliary data used here.  We used images
form the Southern H Alpha Sky Survey Atlas (SHASSA), the {\em Spitzer}, {\em
  IRAS}, {\em ISO}, {\em COBE} and {\em WMAP} satellites, and low frequency
data from the Parkes-MIT-NRAO (PMN) survey, the Rhodes/HartRAO survey, the
Stockert radio-telescope, and the Haslam 408~MHz survey.  Figures
\ref{fig:mosaic_zoph} and \ref{fig:mosaic_LDN1780} show large fields of
3$^{\circ}$ around the two clouds from some of the aforementioned data.

{\small    %%% TABLE 3
  \begin{table}
    \centering
    \caption{Flux densities extracted within a 1$^{\circ}$ circular aperture for $\zeta$~Oph and LDN~1780.}
    \begin{tabular}{lcccc}
      \hline
      \hline
      Freq.             & Telescope
      /                 & $\theta^{1}$ & Flux Density       & Flux Density   \\
      (GHz)             & Survey                   & (~arcmin)  &$\zeta$ Oph (Jy)   &  LDN~1780 (Jy)  \\
      \hline                                                                                      \\
      0.408             & Parkes     & 51        & 55.4 $\pm$ 5.5  & 39.6 $\pm$ 4.0     \\
      1.4               & Stockert   & 40        & 66.9 $\pm$ 6.7  & 56.9 $\pm$ 5.7    \\
      2.3               & HartRAO    & 20        & 19.2 $\pm$ 1.9  & 9.5  $\pm$ 1.0    \\
      2.7               & Stockert   & 20.4      & 23.7 $\pm$ 2.4  & 14.9 $\pm$ 1.4    \\
      23                & {\em WMAP} & 52.8      & 7.6 $\pm$ 0.1   & 1.5 $\pm$ 0.1    \\
      33                & {\em WMAP} & 36.6      & 7.0 $\pm$ 0.2   & 0.9 $\pm$ 0.2    \\
      41                & {\em WMAP} & 30.6      & 7.0 $\pm$ 0.2   & 0.6 $\pm$ 0.2    \\
      61                & {\em WMAP} & 21        & 6.2 $\pm$ 0.4   & $<$ 1.0 (3$\sigma$)\\
      94                & {\em WMAP} & 13.2      & 7.2 $\pm$ 0.6   & $<$ 2.1 (3$\sigma$)\\
      1249              & DIRBE      & 42        & 4940 $\pm$ 580  & 1600 $\pm$ 90       \\       
      2141              & DIRBE      & 42        & 7570 $\pm$ 300  & 4150 $\pm$ 320      \\
      2997              & DIRBE      & 42        & 5600 $\pm$ 200  & 2900 $\pm$ 160     \\
      \hline
    \end{tabular}
    \begin{flushleft}
      $^1$: $\theta$ is the angular resolution FWHM \\
    \end{flushleft}
    \label{tab:sedtable}
  \end{table}
}

%%%%%%%%%%%%%%%%%%%%%%%%%%%%%%%%%%%%%%%%%%%%%%%%%%%%%%%%%%%%%%%%%%%%%%%%%%%%
%%%%%%%%%%%%%%%%%%%%%%%%%%%%%%%%%%%%%%%%%%%%%%%%%%%%%%%%%%%%%%%%%%%%%%%%%%%%

\section{Analysis}

\subsection{Expected radio emission from H${\alpha}$ emission.}
\label{sec:Ha_free}

The radio free-free emission must be accurately known in order to quantify the
contribution of any dust-related excess emission at GHz-frequencies. One way
to estimate its contribution is with the H$\alpha$ emission intensity,
provided that this line is the result of in-situ recombination and not
scattering by dust.

We used the continuum-corrected H$\alpha$ image from the SHASSA survey
\citep{gaustad01}. On the region of $\zeta$~Oph, the image was saturated in
the position of the star $\zeta$~Oph so re-processing was necessary to get a
smooth image.  We masked the saturated pixels and interpolated their values
using the adjacent pixels. Then, a median filter was applied to remove the
field stars.  The processed image was corrected for dust absorption using the
$E(B-V)$ template from \citet{sfd98} and the extinction curve given by
\citet{cardelli89}. The extinction at H${\alpha}$ is $A$(H${\alpha}$) = 0.82
$A$(V) and using $R_V=A(V)/E(B-V)=3.1$ we have that $A$(H${\alpha}$) = 2.54
$E(B-V)$. Finally, we generated a free-free brightness temperature map at
31~GHz using the relationship between H${\alpha}$ intensity and free-free
brightness temperature presented in \citet{DDD} assuming a typical electron
temperature $T_e = 7000$~K, appropriate for the solar neighbourhood.

\subsection{Spectral Energy Distribution fit}

In this analysis we did not use the CBI data because of the large flux loss on
angular scales larger than a few times the synthesised beam. We performed
simulations to estimate the flux loss after the reconstruction and we could
only recover $\sim$20\% of the flux. This would imply that most of the
emission from these two clouds is diffuse with respect to the CBI and CBI2
beam.

The images were smoothed to a common resolution of 53~arcmin, the lowest
resolution of the data used ({\em WMAP}~K). We integrated the auxiliary data
in a circular aperture 1$^{\circ}$ in diameter around the central coordinates
of the 31~GHz data.  Background emission was subtracted integrating in an
adjacent region close to the aperture. This is necessary in the case of the
low frequency radio data because these surveys have large baseline
uncertainties (see for example \citealp{reich09}, or the discussion in
\citealp{davies96}) In Table \ref{tab:sedtable} we present flux densities for
$\zeta$~Oph and LDN~1780 in this aperture.

\subsubsection{$\zeta$ Oph}

\begin{figure} % fig 6
  \centering
  \includegraphics[width=0.45\textwidth]{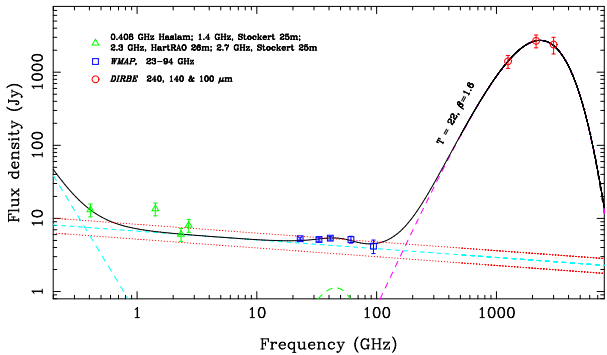}
  \caption{SED of $\zeta$ Oph on 1$^{\circ}$. The dotted red line shows the
    maximum and minimum free-free contribution expected given the H$\alpha$
    emission. In green is the spinning dust emission from a model of
    \citet{ali09}.}
  \label{fig:sed_zoph}
\end{figure}

Fig. \ref{fig:sed_zoph} shows the $\zeta$~Oph SED within a 1$^{\circ}$
aperture, as tabulated in Table \ref{tab:sedtable}. The model is the sum of
synchrotron, free-free, thermal dust and spinning dust emissions. We used a
synchrotron spectral-index ($T_b\propto\nu^{\beta_S}$) with $\beta_S=-2.7$
\citep{davies96}. The free-free power-law is fixed to the predicted brightness
temperature at 31~GHz obtained from the H$\alpha$ image. The spectral index
used is $\beta_{ff}=-0.12$ \citep{DDD}. The dominant source of error in the
determination of the free-free emission through the H$\alpha$ line is the
correction for dust absorption. In Fig. \ref{fig:sed_zoph}, the dotted lines
denote limits for the free-free law: the lower is the emission expected from
the H$\alpha$ image without dust absorption correction and the upper dotted
line is the emission expected with the correction (we assume that all the
absorption occurs as a foreground to the H$\alpha$ emission). A modified black
body with fixed emissivity index $\beta = 1.6$ and 24~K fits the {\em DIRBE}
240, 140 and 100~$\mu$m data points. The spinning dust component was fitted
using the \citet{ali09} models. The main parameters used in the spinning dust
model were $n(H)=100$~cm$^{-3}$ and a temperature of 30~K. The emissivities
are given in terms of the H column density and we used $N_H=1.4\times
10^{21}$cm$^{-2}$ \citep{morton75}.  The parameters of the dust properties are
reasonable for these environments, and were taken from
\citet{WeingartnerDraine01}.

On 1$^{\circ}$ spatial scales, the free-free emission from the H{\sc ii}
region dominates the spectrum in the cm range.  The angular resolution of the
auxiliary radio data is lower than that of the CBI, so we cannot make a flux
comparison on the CBI angular scales.

\subsubsection{Surface brightness spectral comparison}

\begin{figure}  % fig 7
  \centering
  \includegraphics[angle=-90,width=0.45\textwidth]{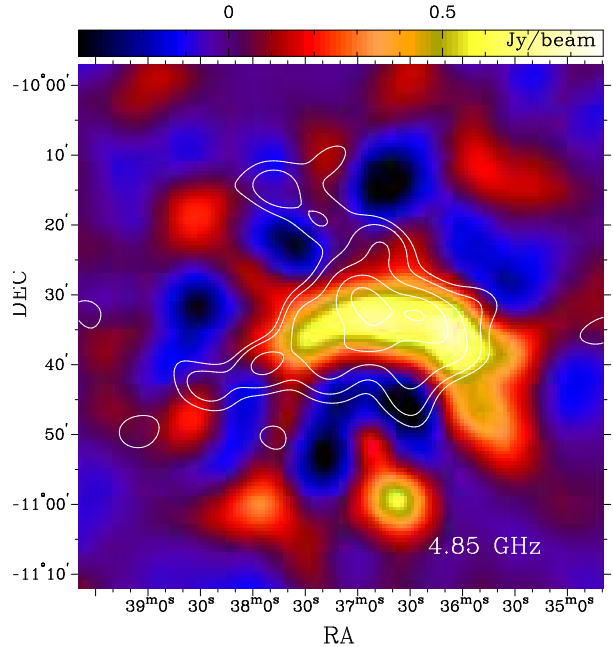}
  \caption{Simulated 4.85~GHz image of $\zeta$~Oph from PMN survey. Contours
    as in Fig. \ref{fig:restored}}
  \label{fig:pmn_zoph}
\end{figure}

The PMN image (Fig. \ref{fig:pmn_zoph}) has an angular resolution similar to
that of the CBI. However this survey has been high-pass filtered; extended
emission on scales larger than $\sim 20$~arcmin is removed \citep{PMN93}. To
make a comparison on CBI angular scales, we simulated visibilities of the PMN
image using the same \uv~coverage of our CBI data. Since large scale modes
have been removed from the PMN data, we exclude visibilities at \uv~radii
$<137$ wavelengths (corresponding to angular scales $\lesssim 25$~arcmin).
Ideally we should consider only visibilities that correspond to angular scales
$\lesssim 15$~arcmin to make a conservative analysis, but doing this excludes
most of the visibilities resulting in no signal. Table
\ref{specific_intensity} lists the surface brightness values at the peak of
the 31~GHz image. The 4.85/31~GHz spectral index is $\alpha_{4.85}^{31}=0.43
\pm 0.20$. We see that $\alpha_{4.85}^{31}$ is significantly different (at the
$2.75\sigma$ level) from $-0.12$, the spectral index if the emission at 31~GHz
were produced by optically thin free-free emission with $T_e \sim
7000$~K. This corresponds to a difference between the 31~GHz intensity and the
value expected from a free-free power-law fixed to the PMN point, the non
free-free specific intensity, of
$I^{\mathrm{nff}}_{31}=I_{31}-I_{4.85}(4.85/31)^{-0.12}= (12 \pm 5)$~mJy
beam$^{-1}$. The significance of the excess is not very high, but we note that
this is a comparison at the peak of the 31~GHz map, which is coincident with
the bulk of the free-free emission.

\begin{table}   %%% TABLE 4
  \centering
  \caption{Surface brightness values in the peak of the 31 GHz image.}
  \begin{tabular}{lccc}
    \hline
    \hline
    Frequency &  Telescope  &  Surface brightness  & Beam size   \\
    (GHz)     &             &      mJy/beam        & (arcmin)    \\
    \hline
    4.85      & Parkes 64m  & 11 $\pm$ 4           &   5.1       \\
    31        & CBI         & 25 $\pm$ 4           &   7.7       \\
    \hline
  \end{tabular}
   \label{specific_intensity}
\end{table}

\subsubsection{LDN~1780}

At the location of LDN~1780, the H$\alpha$ emission present in the SHASSA
image is probably scattered light from the Galactic plane
\citep{mattila07}. Recently, \citet{witt10} confirm this result and show that
the cloud is embedded in a weaker diffuse H$\alpha$ background, of which approximately half is due to scattered light. In contrast, \citet{delBurgo06} state that the H$\alpha$ emission is from
the cloud itself and suggest a very high rate of cosmic rays to explain the
hydrogen ionisation. Whichever is the case, we can set an upper limit to the
free-free contribution using the SHASSA image.  The 100, 140 and 240~$\mu$m
DIRBE and {\em WMAP}~94~GHz points were fitted using a modified black body
with fixed emissivity index $\beta = 2$. The derived temperature is 17~K. For
the spinning dust component we used $n(\text{H})=500$~cm$^{-3}$ and
$T=20$~K. We estimated the H column density using the extinction map from
\citet{sfd98}. This estimate used the relation from \citet{bohlin78} valid for
diffuse clouds:
$N$(H+H$_2$)/$E(B-V)$=5.8$\times10^{21}$cm$^{-2}$mag$^{-1}$. We found
$N_H=3.5\times 10^{21}$cm$^{-2}$.

\begin{figure} % fig 8
  \includegraphics[width=0.45\textwidth]{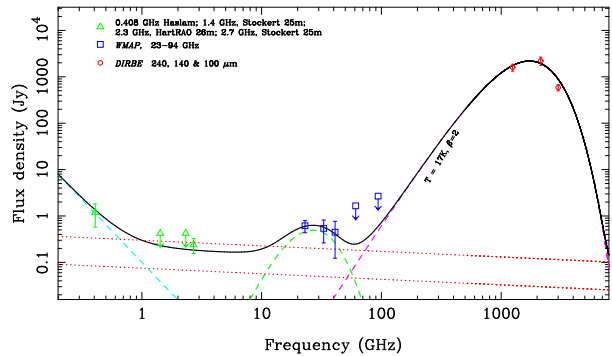}
  \caption{SED of LDN~1780. As in Fig. \ref{fig:sed_zoph}, the red dotted
    lines show upper and lower limits for the free-free contribution. The
    symbols with arrows are upper limits (3~$\sigma$). In green is a spinning
    dust model from a model of \citet{ali09}.  }
  \label{fig:sed_l1780}
\end{figure}

Fig. \ref{fig:sed_l1780} shows the fit. The pair of dotted lines sets limits
on the free-free contribution from the H$\alpha$ data.  On the figure are also
plotted 3~$\sigma$ upper limits to the contribution at 61 and 94~GHz from the
{\em WMAP} V and W bands. This SED shows an excess over optically thin
free-free emission at cm-wavelengths. The spinning dust model is consistent
with the SED in a 1$^{\circ}$ aperture.

\subsection{Morphological analysis}

If dust is responsible for the 31~GHz emission in these clouds, we expect a
morphological correspondence with IR emission. A discussion of the infrared
emission from dust can be found in \citet{DLi07} and references therein. The
100~$\mu$m emission is due to grains bigger than 0.01~$\mu$m that are in
equilibrium with the interstellar radiation field at a temperature $\sim$10-20
K. On the other hand, the mid-IR emission traces VSGs at $\sim$100~K. They are
too hot to be in equilibrium with the environment so these grains are heated
stochastically by starlight photons and, given the very small heat capacity of
a VSG, a single UV photon increases the particle temperature enough to emit at
$\lambda < 60 ~\mu$m.

\begin{figure*}
  \centering % fig 9
  \includegraphics[width=0.28\textwidth,angle=-90]{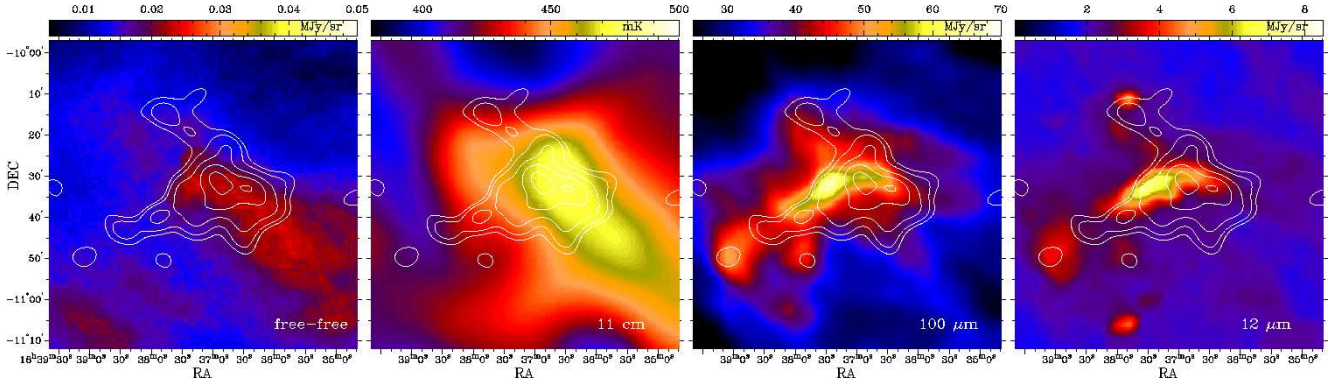}
  \caption{Comparison of different templates of $\zeta$~Oph on the sky plane.
    From the left to right we show the free-free template obtained from the
    H$\alpha$ image, the Stockert 2.7~GHz continuum map, the \iras 100~$\mu$m
    map (which traces large dust grains), and the \iras 12~$\mu$m map (which
    traces very small grains). The contours show the CLEANED image of
    $\zeta$~Oph. Contours are as in Fig.~\ref{fig:restored}. The south-eastern
    ``arm'' of the 31~GHz contours has no counterpart in the free-free
    template; however it resembles the dust emission.}
  \label{fig:zoph_various}
\end{figure*}

In the SED of $\zeta$~Oph the dominant contribution at 31~GHz is free-free
emission. However, inspection of the sky-plane images in
Fig. \ref{fig:zoph_various} suggests that there is no correspondence between
the free-free templates (Fig. \ref{fig:zoph_various} a,b) and the 31~GHz
contours. The CBI data seem to match better with a combination of free-free
and IR emission. We note, however, that the south-eastern arm of the
$\zeta$~Oph cloud is slightly offset by $\sim$~3~arcmin from its IR
counterpart. This offset is larger than the rms pointing accuracy of
CBI, of order 0.5~arcmin.

In LDN~1780 there are clear differences among the IR images
(Fig. \ref{fig:L1780_iras}). By quantifying these differences we can
investigate which kind of dust grains (if big grains or VSGs) are responsible
for the 31~GHz emission in this cloud.

\subsubsection{LDN 1780}

\subsubsection{Sky plane correlations}
\label{sec_sky}

Here we investigate sky-plane cross-correlations.  For this we used MockCBI, a
program which calculates the visibilities $V(u,v)$ of an input image
$I_{\nu}(x,y)$ given a reference \uv~dataset. We computed the visibilities for
the IRIS and free-free templates as if they were observed by the CBI with the
\uv~sampling of our data. We reconstructed these visibilities in the same way
that we did with the CBI data. Finally, we computed the correlation between
the CBI, the IRIS and free-free templates within a square box, 30~arcmin per side,
centred at the phase-centre of the 31~GHz data.

\begin{table}
  
  \caption{ Sky correlation parameters for $\zeta$~Oph. {\em r} is
    the linear correlation coefficient and {\em a} is the 
    proportionality factor between the 31~GHz image and various
    templates in units of $\mu$K~(MJy/sr)$^{-1}$. The errors are 
    given by dispersion in the Monte Carlo simulations.  }
  \centering
  \begin{tabular}{lccccc}
    \hline
    \hline
    & {\em ff} & 12 $\mu$m & 25 $\mu$m & 60 $\mu$m & 100 $\mu$m \\
    \hline
    {\em r}& 0.7$\pm$0.1&0.5$\pm$0.1&0.4$\pm$0.1&0.5$\pm$0.1&0.4$\pm$0.1\\
    {\em a}& 460$\pm$100   & 2.4$\pm$0.5  &2.7$\pm$0.1 &0.2$\pm$0.1& 0.4$\pm$0.1\\
    \hline    
  \end{tabular}  
  \label{tab:sky_corr_zoph}
\end{table}

To estimate the error bars in the correlations, we performed a statistical
analysis. We added Gaussian noise to the observed visibilities and
reconstructed them 1000 times. Then, we correlated the comparison templates
with the noisy mock dataset.

The histogram in Fig. \ref{fig:zoph_hist}a shows the distribution of the
Pearson correlation coefficient of the simulated data. The width of the
distribution gives us an estimate of the uncertainty in the correlation
coefficients and also in the slope of the linear relation between the IR and
31 GHz data.  Fig. \ref{fig:zoph_hist}b shows the distribution of the
proportionality factor between the emission at 31~GHz and the emission at
100~$\mu$m. Table \ref{tab:sky_corr_zoph} lists the derived parameters. The
errors are from the rms dispersion of the Monte Carlo simulations; they are
conservative because of the injection of noise to the data.

\begin{figure*}
  \centering % fig 10
  \includegraphics[width=0.28\textwidth,angle=-90]{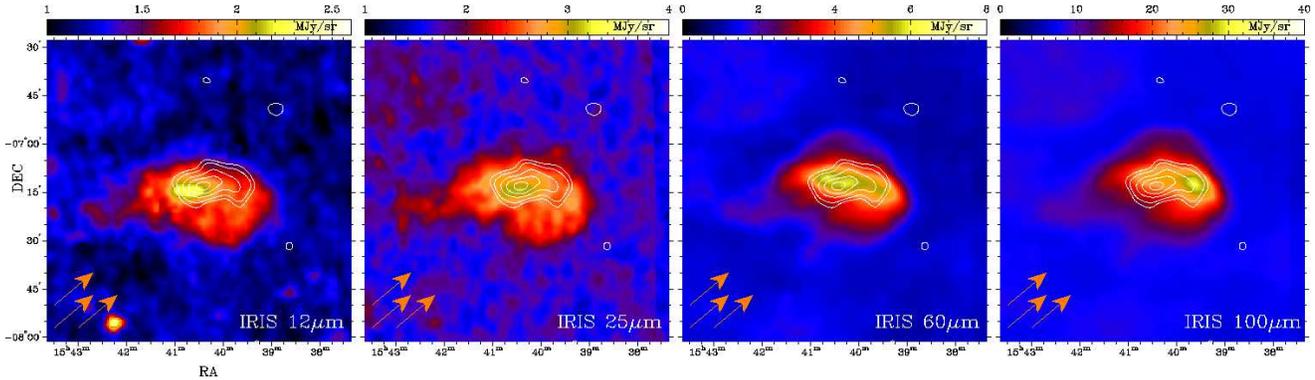}
  \caption{Comparison of the restored CBI2 image of LDN~1780 with {\em IRAS}
    templates. The radio point source NVSS~153909-065843 has been subtracted
    from the CBI2 data. Contours are as in Fig.~\ref{fig:restored}. Note the
    morphological differences between the {\em IRAS} bands. The arrows in the
    corner are perpendicular to the Galactic plane and point towards the north
    Galactic pole.}
  \label{fig:L1780_iras}
\end{figure*}

In the case of $\zeta$~Oph, there are no significant differences between the
different IR data and the best correlation is with the free-free template. As
we could see in the SED of this cloud (Fig. \ref{fig:sed_zoph}), most of the
radio emission is free-free, so it is not a surprise that the best correlation
is with the free-free template.
\begin{figure*}
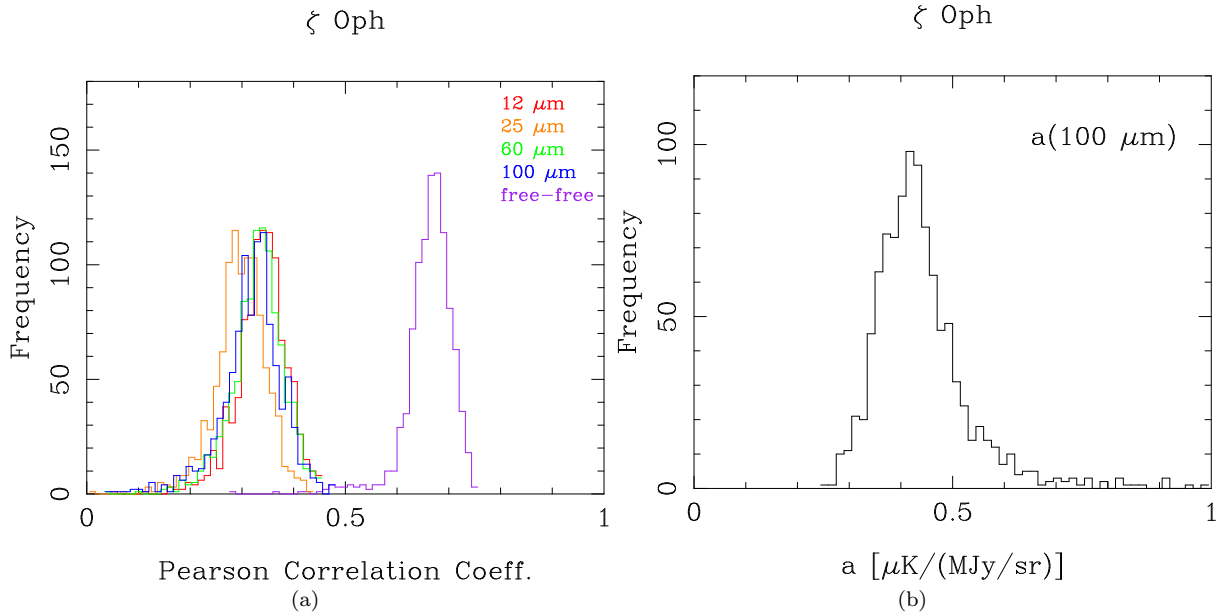
 % fig 11
  \centering
  \subfigure[]{\includegraphics[angle=0,width=0.45\textwidth]{f11_a.eps}}
  \subfigure[]{\includegraphics[angle=0,width=0.45\textwidth]{f11_b.eps}}
  \caption{Histograms form the Monte Carlo simulations.  Left: histogram of
    the distribution of the correlation coefficient $r$ for the different
    templates on $\zeta$~Oph. The correlation with the free-free template is
    significantly better than the correlation with any dust template. Right:
    histogram of the distribution of {\em a}, the proportionality factor
    between the 31~GHz and the 100~$\mu$m images. The rms dispersion of these
    simulations are used as error bars for our results.}
  \label{fig:zoph_hist}
\end{figure*}

\begin{figure*}
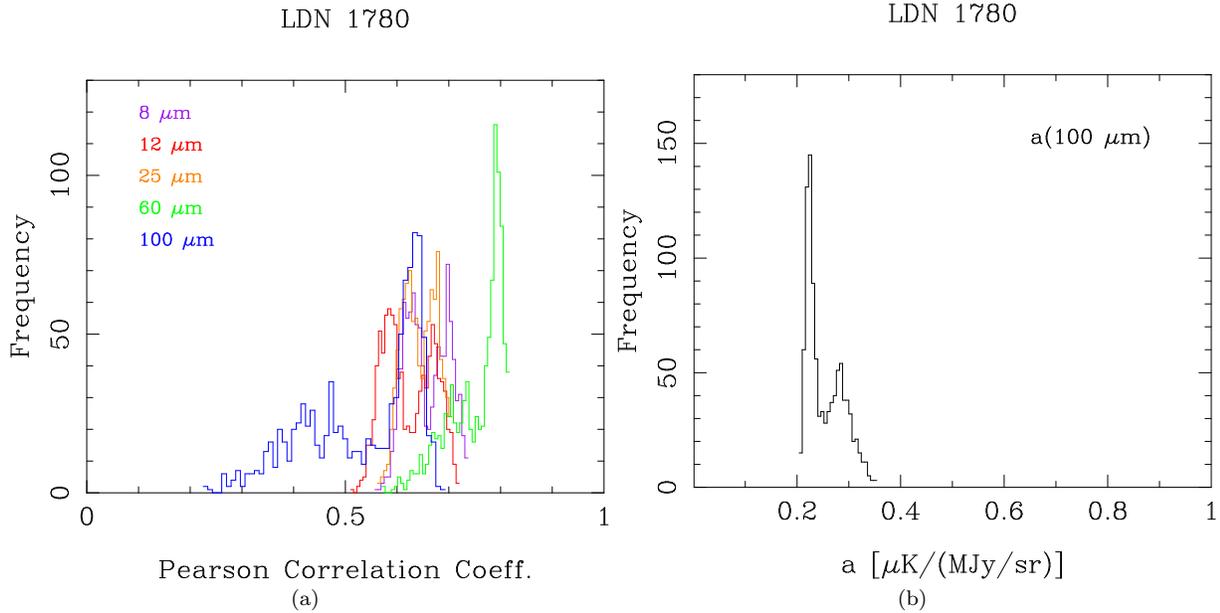
 % fig 12
  \centering
  \subfigure[]{\includegraphics[angle=0,width=0.45\textwidth]{f12_a.eps}}
  \subfigure[]{\includegraphics[angle=0,width=0.45\textwidth]{f12_b.eps}}
  \caption{Histograms form the Monte Carlo simulations for LDN~1780. Left:
    distribution of the correlation coefficient $r$ for the different IR
    templates. Right: histogram of the distribution of {\em a}, the
    proportionality factor between the 31~GHz and the 100~$\mu$m images.}
  \label{fig:l1780_hist}
\end{figure*}

In LDN~1780, the correlations between 31~GHz and the IR data show more
interesting results. Here, all the IR templates correlate better with the CBI
data than in $\zeta$~Oph. The best match is with \iras 60~$\mu$m, as can be
inferred qualitatively from Fig. \ref{fig:L1780_iras}; the Monte Carlo
simulations confirm this result (Fig. \ref{fig:l1780_hist}a). The emission at
60~$\mu$m in the diffuse ISM is mainly from VSGs and a 30-40~\% contribution
from big grains \citep{desert90} so our results favour the idea of a VSGs
origin for the cm-wave radiation. Table \ref{tab:sky_corr_l1780} lists the
results.

PAHs emission at 8 \& 12~$\mu$m have similar correlation coefficients to that
of the 100~$\mu$m emission. However, the emission of VSGs depend on the
strength of the IRF. The differences in morphology that appear in this cloud
depend both on the distribution of grains within the cloud and in the manner
that the cloud is illuminated by the IRF. In the next section we investigate
how the IRF illuminates this cloud.

\begin{figure*}
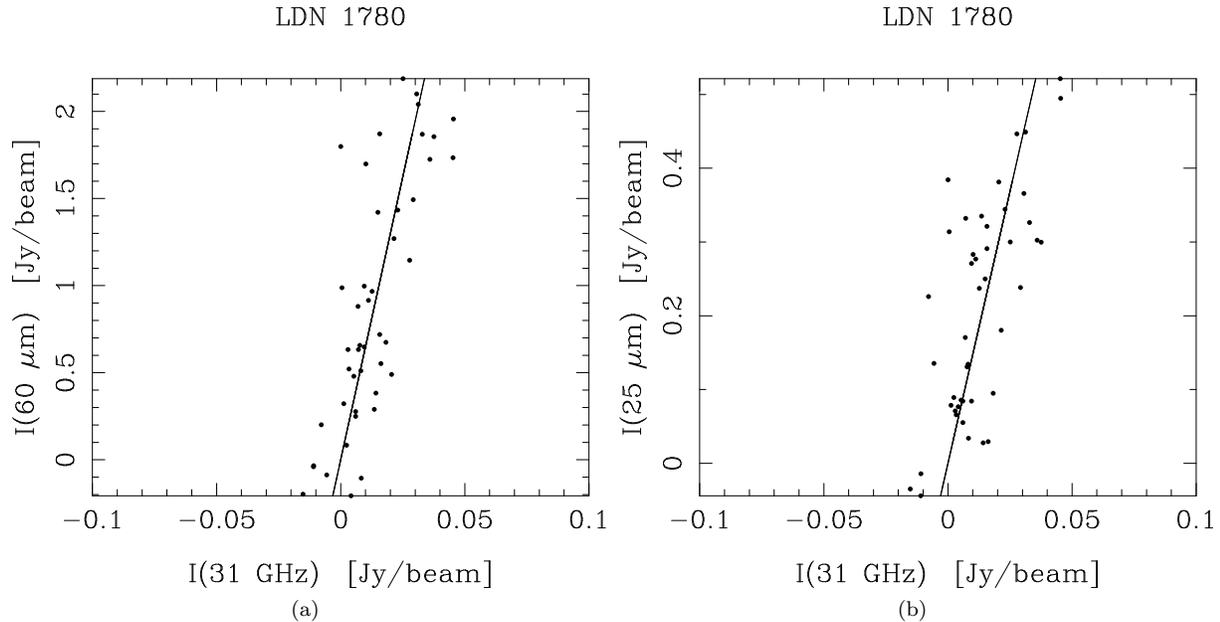
 % fig 13
  \centering
  \subfigure[]{\includegraphics[angle=0,width=0.45\textwidth]{f13_a.eps}}
  \subfigure[]{\includegraphics[angle=0,width=0.45\textwidth]{f14_b.eps}}
  \caption{Pixel-by-pixel correlation of CBI data with \iras 60~$\mu$m and the
    corrected \iras 25~$\mu$m templates on LDN~1780. \iras 60~$\mu$m
    correlates better in the more diffuse regions while IRAC 25~$\mu$m does
    better in the peak of the CBI image.}
  \label{fig:l1780_corr_sky}
\end{figure*}

\begin{table}
  \caption{\label{tab:sky_corr_l1780} Correlation parameters for LDN~1780.
    {\em r} is the linear correlation coefficient and {\em a} is the 
    proportionality factor between the 31~GHz image and various
    templates in units of $\mu$K~(MJy/sr)$^{-1}$. The errors are 
    given by the dispersion in the Monte Carlo simulations.}  
  \centering
  \begin{tabular}{lccccc}
    \hline
    \hline
           & 8 $\mu$m   & 12 $\mu$m & 25 $\mu$m  & 60 $\mu$m & 100 $\mu$m\\
    \hline
    {\em r}& 0.6$\pm$0.1&0.5$\pm$0.1&0.7$\pm$0.05&0.8$\pm$0.1&0.6$\pm$0.1\\
    {\em a}& 5.3$\pm$1.0&5.2$\pm$1.4&3.7$\pm$0.9 &0.9$\pm$0.2&0.2$\pm$0.1\\
    \hline    
    \end{tabular}  
  \end{table}

\subsubsection{IRF on LDN~1780}
\label{sec:irf}
The radio emission from spinning VSGs is fairly independent of the IRF
\citep{DL98b, ali09, ysard10b}.  On the other hand, their near-IR (NIR)
emission is due to stochastic heating by interstellar photons so it is
proportional to the intensity of the radiation field and to the amount of
VSGs.  Therefore, NIR templates corrected for the IRF would trace better the
distribution of VSGs.

The intensity of the radiation field G$_0$ can be estimated \citep[as
in][]{ysard10} from the temperature of the big grains $T_{BG}$ in the
cloud. We constructed a temperature map of those grains fitting a modified
black body to the {\em ISO} 100 \& 200~$\mu$m images pixel-by-pixel (at the
same resolution). With this temperature map, we calculated G$_0$ given that:
\begin{equation}
  \text{G}_0= \left ( \frac{T_{BG}}{17.5~{\text K}}\right )^{\beta+4},
  \label{eq:G0}
\end{equation}
with $\beta=2$. We divided the NIR templates (8 \& 12~$\mu$m) by our G$_0$ map
and then repeated the correlations with the 31~GHz data. Again, to estimate
error bars, we used the Monte Carlo simulations. The correlations in this
case are tighter than with the uncorrected templates; here $r=0.69\pm0.04$, a
2~$\sigma$ improvement.

%%%%%%%%%%%%%%%%%%%%%%%%%%%%%%%%%%%%%%%%%%%%%%%%%%%%%%%%%%%%%%%%%%%%%%%%%%%%
%%%%%%%%%%%%%%%%%%%%%%%%%%%%%%%%%%%%%%%%%%%%%%%%%%%%%%%%%%%%%%%%%%%%%%%%%%%%

\section{Comparison of 31~GHz emissivities}

One motivation for this work is that the physical conditions in translucent
clouds approach that of the cirrus clouds. Here, we compare the radio emission
of different clouds in terms of their $N(H)$. To avoid differences in beam
sizes and frequencies observed, we choose to compare only sources observed by
the CBI, at 31~GHz and scales of 4-8~arcmin.  We also calculate an averaged
column density for the cirrus clouds using the extinction map from
\citet{sfd98} in the positions observed by \citet{leitch97} at frequency of
32~GHz and a beam size $\sim$7~arcmin. This beam and frequency are similar to
those of the CBI.

\begin{figure}
  \centering
  \includegraphics[width=0.45\textwidth, angle=0]{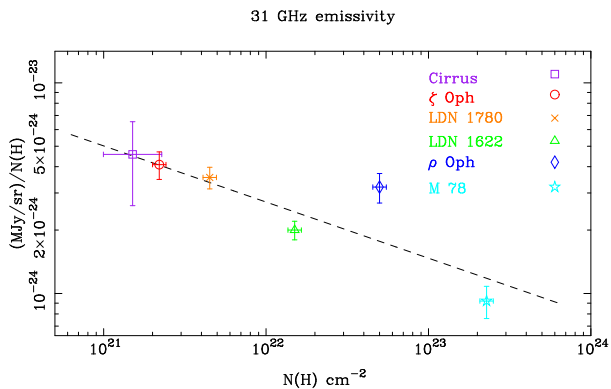}
  \caption{Emissivity at 31~GHz versus column density for different clouds
    observed with the CBI. The point corresponding to cirrus clouds was
    obtained as an average from the results of \citet{leitch97}. The dashed
    line is the best linear fit to the data.}
  \label{fig:emi}
\end{figure}

In Table \ref{emissivities_comparison}, we list the radio intensity at the
peak of the CBI images, alongside a value for the column density at the same
position and the ratio between these two quantities. In Fig.~\ref{fig:emi} we
plot these quantities. The linear fit shown has a slope of $0.54\pm
0.10$~MJy~sr$^{-1}$~cm$^{-4}$ and the correlation coefficient is $r=-0.87
$. There seems to be a trend in the direction of diminishing emissivity with
increasing column density.  Despite the large variations in column density
($\sim$2-3 orders of magnitude), it is interesting that the emissivities of
the clouds lie in a small range of $\sim$1 order of magnitude. However, it is
worth noting that {\em if} the inverse relation we see is real, it will
indicate that the anomalous emission is not associated with large dust grains,
since their number increases with density, because of dust growth.  A similar
result was obtained by \citet{lagache03}, who used {\em WMAP} data combined
with IR templates and gas tracers in the whole sky on angular scales of 7~deg.

\begin{table} % table 7
  \caption{Emission parameters for different clouds, all of them but
    the cirrus were observed by the CBI at 31~GHz. The second column is the
    column density, the third is the peak intensity at 31~GHz and the fourth
    is the emissivity at 31~GHz. References are the following:
    (1) \citet{leitch97}, (2) \citet{casassus1622}, (3) \citet{casassus08}, (4)
    \citet{castellanos10}
  }
  \centering
  \begin{tabular}{lcccc}
    \hline
    &           $N(H)$~$^a$ &   $I_{31}$~$^b$     & $\epsilon_{31}$~$^c$ \\
    \hline
    Cirrus$^1$    & $0.15\pm0.07$  &   $6.9\pm1$    &$4.6\pm2.0$ \\
    $\zeta$~Oph   & $0.22\pm0.02$  &   $9\pm1$      &$4.1\pm0.6$ \\    
    LDN~1780      & $0.45\pm0.04$  &   $16\pm1$     &$3.5\pm0.4$ \\
    LDN~1622$^2$  & $1.5\pm0.15$   &   $30\pm2$     &$2.0\pm0.2$ \\
    $\rho$~Oph$^3$& $5.0\pm0.50$   &   $180\pm20$   &$3.2\pm0.5$ \\
    M78$^4$       & $22.8\pm0.23$  &   $210\pm30$   &$0.9\pm0.1$ \\
    \hline
  \end{tabular}
  \label{emissivities_comparison}
  \begin{flushleft}
    $^a$: $\times10^{22}$~cm$^{-2}$ \\
    $^b$: $\times10^{-3}$~MJy~sr$^{-1}$ \\
    $^c$: $\times10^{-24}$~MJy~sr$^{-1}$~cm$^{2}$
  \end{flushleft}
\end{table}

\section{Conclusions}
We have presented 31~GHz data of two translucent clouds, $\zeta$~Oph and LDN
1780 with the aim of characterising their radio emissivities. We found an
anomalous emission excess in both clouds at 31~GHz on angular scales of
$\sim$7~arcmin in $\zeta$~Oph and $\sim$5~arcmin in LDN~1780.

The SED of $\zeta$ Oph on large (1$^{\circ}$) angular scales is
dominated by free-free emission from the associated H{\sc ii}. Because of
this, it is difficult to quantify the contribution from dust to the 31~GHz
data. However, when comparing with the optically thin free-free extrapolated
from the 5~GHz PMN image, we find a 2.4~$\sigma$ excess at 31~GHz on spatial
scales of 7~arcmin in surface brightness.

In the SED of LDN~1780 we see an excess on 1$^{\circ}$ angular scales. The
free-free contribution in this cloud is expected to be very low; the H$\alpha$
emission may be scattered light from the IRF, so it would not have a radio
counterpart. A spinning dust component can explain the anomalous emission
excess in the SED. Correlations between the cm-wave data and IR-templates
shows a trend favouring \iras 25 \& 60~$\mu$m. The best match in this case is
with \iras 60~$\mu$m although the peak of the CBI image is best matched by
IRAC 8~$\mu$m. We corrected the IRAC 8~$\mu$m and \iras 12~$\mu$m by the IRF
and found a tighter correlation with these corrected templates.  In the
spinning dust models, the VSGs dominate the radio emission. Our results
support this mechanism as the origin for the anomalous emission in this cloud.

The 31~GHz emissivities found in both clouds are similar and have intermediate
values between the cirrus clouds and the dark clouds. The emissivity
variations with column density are small, although we find an indication of an
inverse relationship, which would further support a VSGs origin for the
cm-emission.  The anomalous foreground which contaminates CMB data comes from
cirrus clouds at typically high Galactic latitudes. Here, we see that there is
not a large difference between the radio emissivity of cirrus and translucent
clouds on 7~arcmin angular scales. Because of this similarity, translucent
clouds are good places to investigate the anomalous CMB foreground.

\section*{Acknowledgments}
We are most grateful to Mika Juvela who kindly shared with us the LDN~1780
{\em ISO} images. We thank an anonymous referee for a thorough reading and
very useful comments. MV acknowledges the funding from Becas Chile.
S.C. acknowledges support from a Marie Curie International Incoming Fellowship
(REA-236176), from FONDECYT grant 1100221, and from the Chilean Center for
Astrophysics FONDAP 15010003. CD acknowledges an STFC Advanced Fellowship and
ERC grant under FP7. LB and RB acknowledge support from CONICYT project Basal
PFB-06. This work has been carried out within the framework of a NASA/ADP
ROSES-2009 grant, n. 09-ADP09-0059. The CBI was supported by NSF grants
9802989, 0098734 and 0206416, and a Royal Society Small Research Grant. We are
particularly indebted to the engineers who maintained and operated the CBI:
Crist\'{o}bal Achermann, Jos\'{e} Cort\'{e}s, Crist\'{o}bal Jara, Nolberto
Oyarace, Martin Shepherd and Carlos Verdugo.  This work was supported by the
Strategic Alliance for the Implementation of New Technologies (SAINT - see
www.astro.caltech.edu/chajnantor/saint/index.html) and we are most grateful to
the SAINT partners for their strong support. We gratefully acknowledge support
from the Kavli Operating Institute and thank B. Rawn and S. Rawn Jr.  We
acknowledge the use of the Legacy Archive for Microwave Background Data
Analysis (LAMBDA). Support for LAMBDA is provided by the NASA Office of Space
Science. We used data from the Southern H-Alpha Sky Survey Atlas (SHASSA),
which is supported by the National Science Foundation.

\bibliographystyle{mn2e}

\label{lastpage}

\end{document}